\documentclass[reprint, prl, aps, twocolumn, showkeys]{revtex4-1}

\usepackage{amssymb}
\usepackage{amsmath}
\usepackage{esvect}
\usepackage{graphicx}
\usepackage{bm}
\usepackage{dcolumn}
\usepackage{color}
\usepackage{ulem}
\usepackage{float}
\usepackage{textcomp}
\restylefloat{table}
\usepackage{verbatim}
\usepackage{epstopdf}
\usepackage[colorlinks]{hyperref}
\usepackage[mediumspace,mediumqspace,squaren]{SIunits}

\begin{document}

\title{Anomalous coupling between magnetic and nematic orders in quantum Hall systems}
\date{\today}

\author{Md.\ Shafayat Hossain}
\author{M. A.\ Mueed}
\author{Meng K.\ Ma}
\author{Y. J.\ Chung}
\author{L. N.\ Pfeiffer} 
\author{K. W.\ West}
\author{K. W.\ Baldwin}
\author{M.\ Shayegan}
\affiliation{Department of Electrical Engineering, Princeton University, Princeton, New Jersey 08544, USA}

\begin{abstract}
The interplay between different orders is of fundamental importance in physics. The spontaneous, symmetry-breaking charge order, responsible for the stripe or the nematic phase, has been of great interest in many contexts where strong correlations are present, such as high-temperature superconductivity and quantum Hall effect. In this article we show the unexpected result that in an interacting two-dimensional electron system, the robustness of the nematic phase, which represents an order in the charge degree of freedom, not only depends on the orbital index of the topmost, half-filled Landau level, but it is also strongly correlated with the magnetic order of the system. Intriguingly, when the system is fully magnetized, the nematic phase is particularly robust and persists to much higher temperatures compared to the nematic phases observed previously in quantum Hall systems. Our results give fundamental new insight into the role of magnetization in stabilizing the nematic phase, while also providing a new knob with which it can be effectively tuned.
\end{abstract} 

\maketitle

Besides the isotropic phases of nature such as the fractional quantum Hall and composite fermion liquids, interacting fermionic systems host a tantalizing nematic phase where the charges spontaneously organize into symmetry-breaking, stripe-like clusters  \cite{Fradkin.ARCMP.2010, Shayegan.Flatland.Review.2006}. Its existence was first predicted for a two-dimensional electron system (2DES) subjected to a perpendicular magnetic field, where the electronic kinetic energy  is quantized into a set of Landau levels (LLs). When a LL with high orbital index is half filled, in the presence of long-range Coulomb interaction a spontaneous stripe-like charge ordering can emerge, manifesting periodic density (or LL filling-factor) oscillations along one spatial direction \cite{Koulakov.PRL.1996, Moessner.PRB.1996, Fogler.PRB.1996}. Quantum and thermal fluctuations, as well as disorder, affect the strictly periodic nature of these oscillations and induce a nematic order \cite{Fradkin.ARCMP.2010, Fradkin.PRB.1999}. Soon after the theoretical predictions, experimental signatures of the nematic phases were reported as strong anisotropies in the two in-plane transport directions (higher resistance along the direction of charge oscillations), in very high-mobility 2DESs \cite{Lilly1.PRL.1999, Du.SSC.1999} and 2D hole systems  \cite{Shayegan.PhysicaE.2000} confined to GaAs quantum wells (QWs). These were followed by reports of nematic phases in a variety of bulk systems such as Sr$_{3}$Ru$_{2}$O$_{7}$ \cite{Borzi.Science.2007} and high-temperature superconductors \cite{Ando.PRL.2002,Hinkov.Science.2008, Kuo.Science.2016}. Such ubiquity raises the question whether nematic ordering competes or is intertwined with magnetism, high-temperature superconductivity, quantum Hall effect, and quantum criticality. Although the existence of nematicity and some of its macroscopic properties have been scrutinized \cite{Fradkin.ARCMP.2010, Shayegan.Flatland.Review.2006, Koulakov.PRL.1996, Moessner.PRB.1996, Fogler.PRB.1996, Fradkin.PRB.1999, Lilly1.PRL.1999, Du.SSC.1999, Shayegan.PhysicaE.2000, Pan.PRL.1999, Lilly.PRL.1999, Jungwirth.PRB.1999, Kukushkin.PRL.2011, Liu.PRB.2013, Liu.PRB.2016, Samkharadze.Natphys.2016, Friess.Natphys.2017, Qian.Natcom.2017, Shi.PRB.2017, Shi.PRB.2015, Borzi.Science.2007, Ando.PRL.2002, Hinkov.Science.2008, Kuo.Science.2016, Wexler.IJMP.2006, footnote1, Zeitler.PRL.2001, Pan.PRB.2001, Chalker.PRB.2002}, understanding the interplay between the nematic and other intricate charge or magnetic orders remains a challenging problem in condensed matter physics \cite{Fradkin.ARCMP.2010, Fernandes.Natphys.2014}.

Here we unravel an unexpected coupling between magnetic and nematic orders at half-filled LLs in a novel 2DES confined to an AlAs QW that enables a complete tuning of magnetization. Thanks to the comparable magnitudes of the cyclotron and Zeeman energies in this 2DES, tilting the sample in the magnetic field allows us to tune the Fermi energy between LLs with different orbital ($N$) and spin ($\uparrow$ and $\downarrow$) indices, and capture a complete evolution of the ground states as a function of $N$ and the magnetization of the 2DES. For a half-filled LL with $N=0$, there is no nematic phase, regardless of the magnetization. For $N>0$ LLs, when the magnetization of the topmost, half-filled LL is opposite to the magnetization of the underlying LLs, the nematic phase is absent, but when it is aligned a nematic phase is seen. If the half-filled and the underlying LLs are fully magnetized, the nematic phase is anomalously robust and persists at temperatures as high as $\simeq2$ K. Our data provide a fresh outlook at the nematic phases in 2DESs as they highlight the importance of the spin degree of freedom, a factor that has been seldom accessible in previous experiments, and ignored in theoretical studies.

\begin{figure}[t!]
\includegraphics[width=0.47\textwidth]{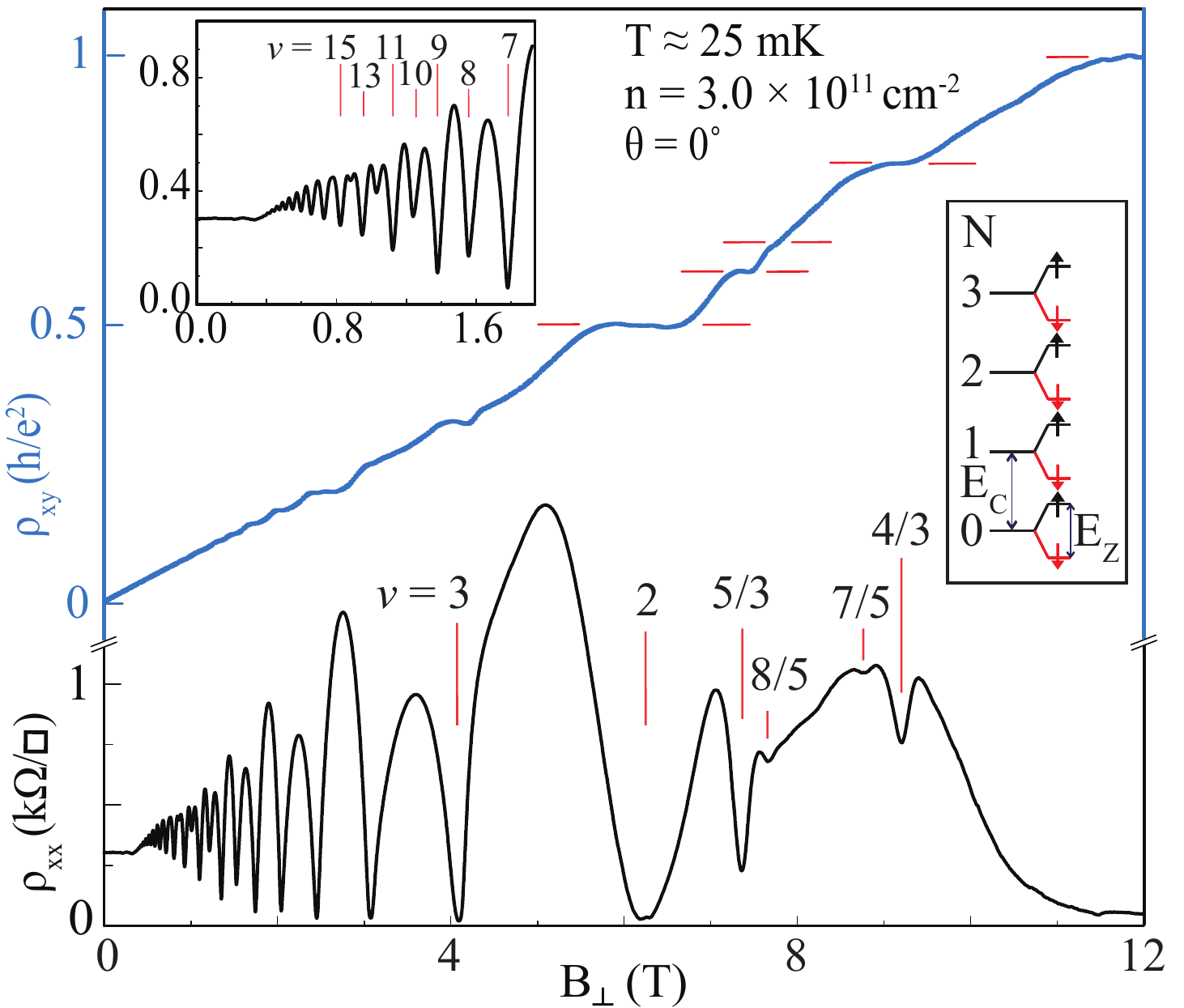}
\caption{\label{fig:Fig1} Magnetotransport traces for our narrow AlAs QW, highlighting Shubnikov-de Haas oscillations at low magnetic fields (upper inset), and integer and fractional quantum Hall effects at higher fields. The right inset is the LL diagram for our 2DES showing the cyclotron ($E_{C}$) and the Zeeman ($E_{Z}$) energies. The red and blue arrows indicate the down-spin and up-spin energy levels for a given LL with orbital index $N$.
}
\end{figure}

\begin{figure*}[t!]
\includegraphics[width=0.99\textwidth]{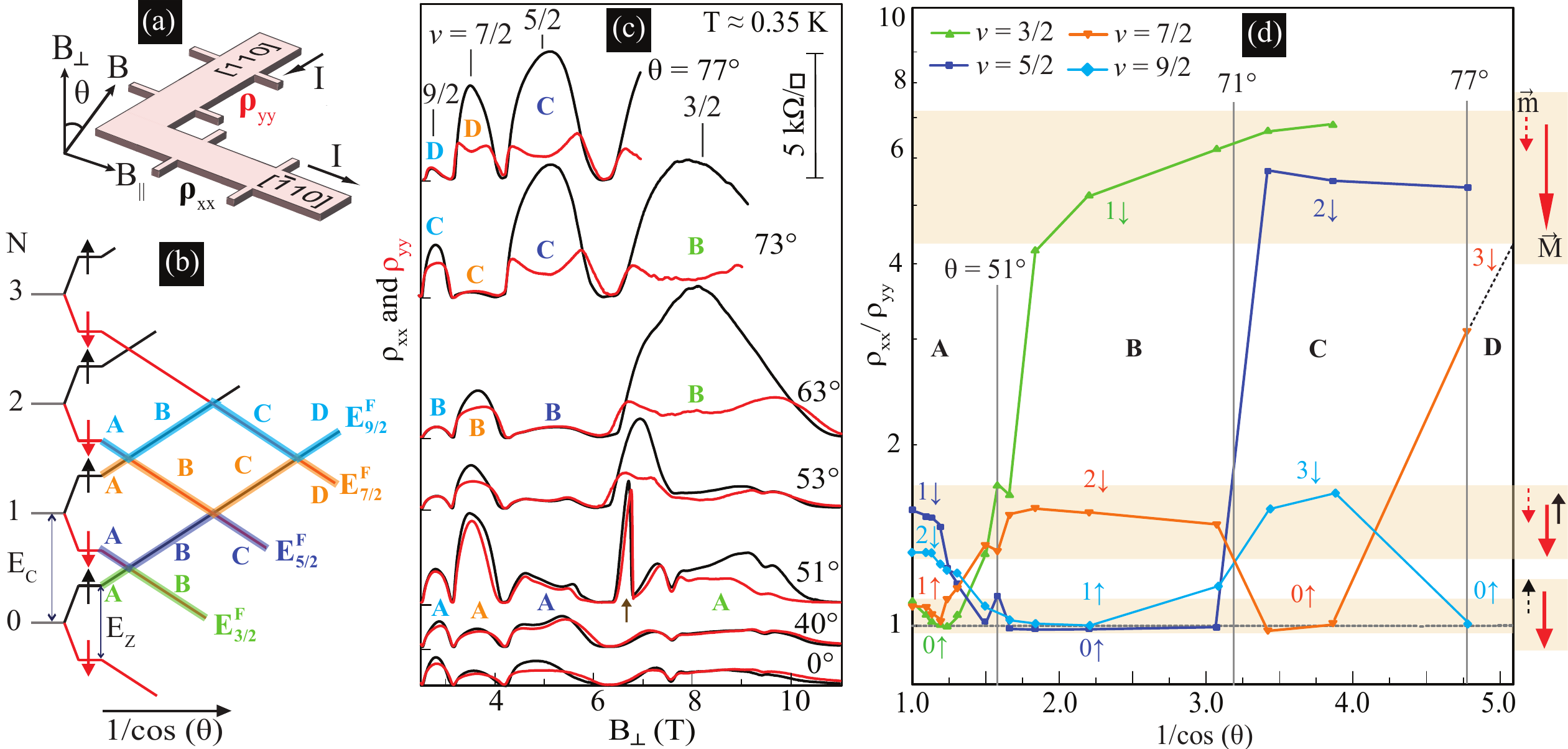}
\caption{\label{fig:Fig2} Evolution of the ground states in the half-filled LLs with tilt. (a) Schematic of our L-shaped Hall bar geometry to measure $\rho_{xx}$ (along $[\overline{1}10]$) and $\rho_{yy}$ (along $[110]$) simultaneously. The in-plane field ($B_{||}$) is parallel to the direction of $[\overline{1}10]$. (b) LL diagram showing the crossing between LLs of opposite spins with tilt. The thick green, dark-blue, orange, and light-blue lines represent the evolution of $E^F$ for $\nu = 3/2$, $5/2$, $7/2$ and $9/2$, respectively. (c) $\rho_{xx}$ and $\rho_{yy}$ at different $\theta$. At $\theta=51^{o}$, the 0$\uparrow$ and 1$\downarrow$ LLs cross, as manifested by the resistance spike at $\nu =2$ marked by a brown arrow. The color-coded A, B, C, D represent the position of $E^F$ shown in (b). (d) Resistance anisotropy vs $1/cos(\theta)$. Large anisotropy is observed when $E_{3/2}^F$, $E_{5/2}^F$, and $E_{7/2}^F$ are in 1$\downarrow$, 2$\downarrow$, and 3$\downarrow$, respectively, and the 2DES is fully magnetized. Vertical grey lines mark the angles of coincidence: $51^{o}$ for the crossing between 1$\downarrow$ and 0$\uparrow$ when $E_{Z} = E_{C}$, $71^{o}$ for the crossing between 2$\downarrow$ and 0$\uparrow$ when $E_{Z} = 2E_{C}$, and $77^{o}$ for the crossing between 3$\downarrow$ and 0$\uparrow$ when $E_{Z} = 3E_{C}$. The arrows to the right of (d) represent the magnetization of the topmost, half-filled LL ($\vec{m}$, dotted arrow) and the underlying occupied LLs ($\vec{M}$, solid arrows) for different shaded bands.
}
\end{figure*}

\begin{figure}[t!]
\includegraphics[width=.42\textwidth]{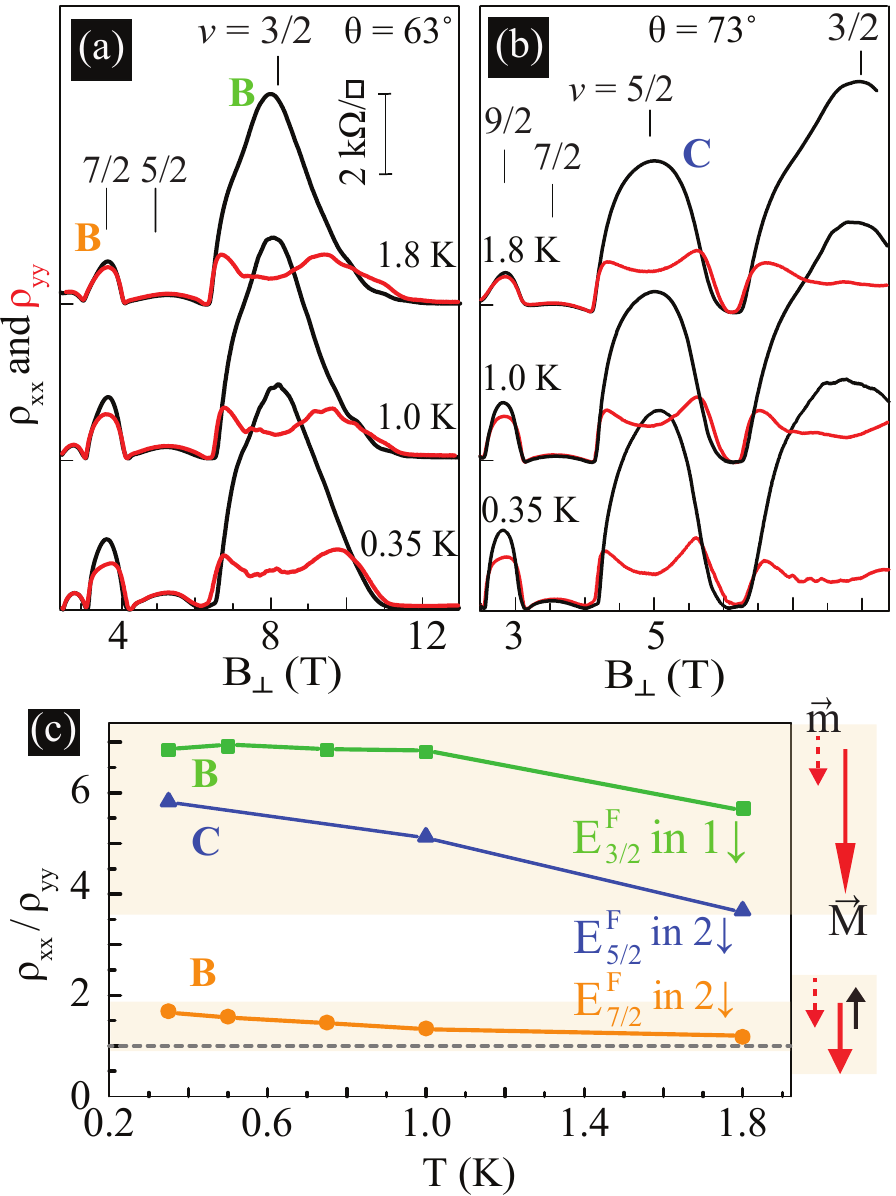}
\caption{\label{fig:Fig3} Temperature dependence of the anisotropic phases. (a) At $\theta = 63^{o}$, $E_{3/2}^F$ is in 1$\downarrow$ and the 2DES is fully magnetized. The transport anisotropy persists even at $T=1.8$ K. A similar robustness is seen in panel (b) at $\theta = 73^{o}$ for $\nu = 5/2$ in 2$\downarrow$ when the 2DES is fully magnetized. On the other hand, when the 2DES is \textit{not} fully magnetized, e.g., when $E_{7/2}^F$ is in 2$\downarrow$ at $\theta = 63^{o}$ (panel (a)) or $E_{9/2}^F$ is in 3$\downarrow$ at $\theta = 73^{o}$ (panel (b)), the anisotropies are small and essentially disappear at $T=1.8$ K. (c) Transport anisotropy vs temperature, indicating robust anisotropy for $\nu = 3/2$ in 1$\downarrow$ and $\nu = 5/2$ in  2$\downarrow$, but weak anisotropy for $\nu = 7/2$ in 2$\downarrow$.}
\end{figure}

\begin{figure}[b!]
\includegraphics[width=0.48\textwidth]{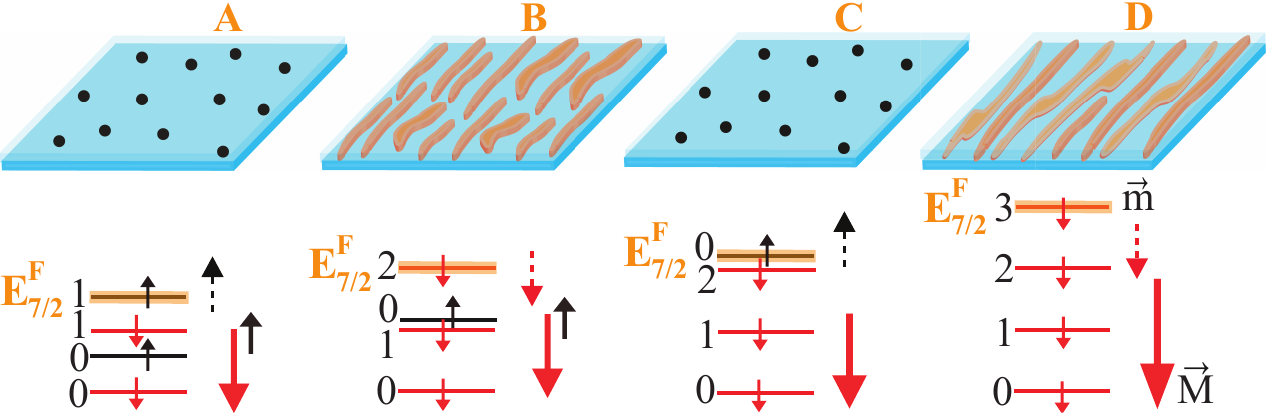}
\caption{\label{fig:Fig4} Illustration of magnetization-driven transitions between Fermi liquid and nematic phases as $E_{7/2}^F$ moves between LLs of different orbital and spin indices, from \textbf{A} (1$\uparrow$, Fermi liquid) to \textbf{B} (2$\downarrow$, weakly nematic), to \textbf{C} (0$\uparrow$, Fermi liquid), and finally to \textbf{D} (3$\downarrow$, strongly nematic).
}
\end{figure}

The 2DES confined in our sample is confined to a very narrow (5.66-nm-thick) AlAs QW and occupies an out-of-plane conduction-band valley with an isotopic in-plane mass \cite{Suppl.Mat.}. It exhibits fractional quantum Hall states (FQHSs) at high magnetic fields in the $N=0$ LL, as seen in Fig. 1. The longitudinal resistivity ($\rho_{xx}$) trace shows strong minima at $\nu=5/3$ and $4/3$, accompanied by corresponding plateaus in the Hall resistivity ($\rho_{xy}$). There are also weak $\rho_{xx}$ minima at $\nu=8/5$ and $7/5$, hinting at developing FQHSs. Figure 1 traces provide evidence for the first observation of FQHSs in an AlAs QW with out-of-plane valley occupation, and attest to the sufficiently high quality of the sample to support many-body states. The 2DES has a relatively large effective Land\'e \textit{g}-factor ($g^*$)and effective mass ($m^*$) \cite{Suppl.Mat.}. These lead to a ratio of $0.63$ for the Zeeman energy ($E_{Z} = g^{*}\mu_{B}B$) and cyclotron energy ($E_{C} = \hbar  eB_{\perp}/m^{*}$) in a purely perpendicular magnetic field, as the LL energy diagram in the right inset to Fig. 1 illustrates. The LL diagram implies that the energy gaps at odd LL fillings ($\nu$) are larger than those at even $\nu$, consistent with the stronger $\rho_{xx}$ minima seen at odd $\nu$ in the low-field Shubnikov-de Haas oscillations (Fig. 1 upper inset). This pattern is in contrast to the one seen in GaAs 2DESs where $E_Z$ is much smaller than $E_C$. More importantly, the close magnitudes of $E_Z$ and $E_C$ allow us to study of the ground states of the 2DES at a given filling as we tune the Fermi level ($E^F$) through different LLs by tilting the sample in the magnetic field; such tuning is extremely challenging in GaAs 2DESs where $E_Z<<E_C$.

To perform the experiments, the sample is mounted on a stage which can be rotated \textit{in situ} (Fig. 2(a)). Since $E_C$ depends on the perpendicular field ($B_{\bot}$) whereas $E_Z$ depends on the total field ($B$), by tilting the sample in field, the LLs of different orbital indices cross as the sample is tilted, as depicted in the LL diagram shown in Fig. 2(b). When two LLs of opposite spins cross, ferromagnetic domains form for each spin, causing an extra dissipation at their boundaries, which manifests as a resistance spike in $\rho_{xx}$ \cite{DePoortere.Science.2000}. An example of such a crossing and resistance spike is seen in the traces taken at $\theta =51^{o}$ (see the vertical brown arrow in Fig. 2(c)), allowing us to pinpoint the first LL crossing. Quantitatively, the crossings occur at angles $\theta_j$ according to the expression $g^*m^*/2m_0 = jcos(\theta_j)$, where $j=$ 1, 2, 3, ..., and $m_0$ is the free electron mass \cite{Vakili.PRL.2004}. From our data, we find $g^*m^*/2m_0= 0.63$, consistent with a previous report for a 2DES confined to a narrow AlAs QW \cite{Vakili.PRL.2004, Shayegan.AlAs.Review.2006}.

In order to probe the anisotropic ground states of the 2DES, we fabricated an L-shaped Hall bar (Fig. 2(a)) for simultaneous measurements of $\rho_{xx}$ and $\rho_{yy}$, the longitudinal resistivities parallel and perpendicular to the direction of the parallel magnetic field, $B_{||}$. The $\rho_{xx}$ and $\rho_{yy}$ data in Figs. 2(c) and 2(d) capture our main findings. They show that the 2DES at half-filled LLs exhibits a remarkable evolution, alternating between isotropic and anisotropic states with different degrees of anisotropy, as the sample is tilted in magnetic field. The measured anisotropy ratios, $\rho_{xx}/\rho_{yy}$, are summarized in Fig. 2(d) as a function of $1/cos(\theta)$; this figure also shows $\theta$ at which different LL coincidences occur (see Fig. 2(b)). 

The seemingly complex evolution observed in Fig. 2(d) can be explained based on three simple rules by considering the orbital index $N$ and the magnetization of the LL where $E^F$ lies at a given $\nu$ and $\theta$ relative to the net magnetization of the underlying LLs: ($i$) \textit{Transport is isotropic when $E^F$ lies in an $N=0$ LL, regardless of the magnetization.} Examples are: $\nu=3/2$ at $\theta = 0^{o}$ ($E_{3/2}^F$ in 0$\uparrow$),  $\nu = 5/2$ in range B ($E_{5/2}^F$ in 0$\uparrow$), and $\nu = 9/2$ at the highest angle as $E_{9/2}^F$ enters 0$\uparrow$. ($ii$) \textit{Transport remains isotropic when $E^F$ lies in an $N>0$ LL as long as the magnetization of the topmost LL is $\uparrow$, i.e., opposite to the net magnetization} ($\downarrow$). Examples are: $\nu=7/2$ at $\theta=0^{o}$ ($E_{7/2}^F$ in 1$\uparrow$), or $\nu = 9/2$ in range B ($E_{9/2}^F$ in 1$\uparrow$). ($iii$) \textit{Transport becomes anisotropic for $N>0$ if the magnetization of the topmost LL is $\downarrow$ so that it aligns with the net magnetization; moreover, the anisotropy is largest once the 2DES becomes fully magnetized, i.e., all the occupied LLs have $\downarrow$ spins.}  Examples for the partially magnetized cases are: $\nu=5/2$ and $\nu=9/2$ at $\theta = 0^{o}$ when the respective $E^F$ lie in 1$\downarrow$ and 2$\downarrow$, $\nu=7/2$ in range B ($E_{7/2}^F$ in 2$\downarrow$), and $\nu=9/2$ in range C ($E_{9/2}^F$ in 3$\downarrow$). In these cases $\rho_{xx}/\rho_{yy}\simeq 1.5$. For the fully magnetized case, examples are $\nu=3/2$ past $\theta=51^{o}$ ($E_{3/2}^F$ in 1$\downarrow$), and $\nu = 5/2$ past $\theta=71^{o}$ ($E_{5/2}^F$ in 2$\downarrow$); in these cases the anisotropy ratio is the largest and can reach $\simeq 7$.

In Fig. 3 we show the evolution of the transport anisotropy with temperature up to $T=1.8$ K. Overall, the magnitude of anisotropy diminishes with increasing T. The data also corroborate the observations summarized in the last paragraph. Namely, when the 2DES is fully magnetized e.g., when $E_{3/2}^F$ lies in 1$\downarrow$ or when $E_{5/2}^F$ resides in 2$\downarrow$, the strong anisotropy persists even at $T=1.8$ K. In contrast, when the 2DES is only partially magnetized and the anisotropy is weak, e.g., $E_{7/2}^F$ lies in 2$\downarrow$, or when $E_{9/2}^F$ resides in 3$\downarrow$ (see Fig. 3(b)), the anisotropy is all but gone at the highest $T$.

Several features of the data presented in Figs. 2 and 3 are qualitatively reminiscent of the transport anisotropies reported at half-filled LLs in very high quality GaAs 2DESs, and are typically interpreted as stripe (or nematic) phases \cite{Fradkin.ARCMP.2010, Shayegan.Flatland.Review.2006, Koulakov.PRL.1996, Moessner.PRB.1996, Fogler.PRB.1996, Fradkin.PRB.1999, Lilly1.PRL.1999, Du.SSC.1999, Shayegan.PhysicaE.2000, Pan.PRL.1999, Lilly.PRL.1999, Jungwirth.PRB.1999, Kukushkin.PRL.2011, Liu.PRB.2013, Liu.PRB.2016, Samkharadze.Natphys.2016, Friess.Natphys.2017, Qian.Natcom.2017, Shi.PRB.2017}. First, the anisotropies are only observed in higher LLs ($N>0$). When $\theta=0^o$, the resistivity along $[\overline{1}10]$ is larger than along $[110]$ at a half-filled LL with high orbital index. Second, when $B_{||}$ is applied along $[\overline{1}10]$ (Fig. 2), or along $[110]$ \cite{Suppl.Mat.}, the resistivity along $B_{||}$ is larger than in the perpendicular direction ($\rho_{xx}/\rho_{yy}>1$), meaning that the ``hard" axis for the nematic phase is along $B_{||}$. Third, for the cases where we observe a strong anisotropy, e.g., when $E_{3/2}^F$ lies in 1$\downarrow$ (Fig. 3(a)) or when $E_{5/2}^F$ is in 2$\downarrow$ (Fig. 3(b)), $\rho_{xx}$ exhibits a maximum while $\rho_{yy}$ shows a minimum. (We also note that a \textit{resistivity} anisotropy of $\simeq 7$ in our Hall bar sample translates to a \textit{resistance} anisotropy factor of $\simeq 60$ in the van der Pauw geometry \cite{Simon.PRL.1999} which is often used to probe the nematic phases in 2DESs \cite{Lilly1.PRL.1999, Du.SSC.1999, Shayegan.PhysicaE.2000, Pan.PRL.1999, Lilly.PRL.1999, Liu.PRB.2013, Liu.PRB.2016, Samkharadze.Natphys.2016, Friess.Natphys.2017, Qian.Natcom.2017, Shi.PRB.2017, Shi.PRB.2015}). These similarities strongly suggest that the anisotropic phases that we observe are signatures of nematic phases \cite{footnote2}.

The data of Figs. 2 and 3 provide very strong evidence that both the degree of anisotropy, as well as its persistence to higher temperatures, are directly linked to the degree of magnetization of the 2DES. They imply that the magnetization of the 2DES helps the formation of a nematic phase at half-filled LLs, and that the nematic phase is most robust when the 2DES is fully magnetized (ferromagnet). We highlight this connection in Figs. 2(d) and 3(c) (shaded bands), and Fig. 4. While the role of magnetism in stabilizing nematic phases has been debated in high-temperature superconductors \cite{Fradkin.ARCMP.2010, Fernandes.Natphys.2014}, it has not been reported for nematic phases in 2DESs at half-filled LLs. A main reason is that such control of magnetization is not achievable in systems with small $g^{*}m^{*}$ such as GaAs 2DESs, because of the enormous $B_{||}$ that would be required.

Returning to the data of Figs. 2 and 3, we make another remarkable observation: $\rho_{xx}$ and $\rho_{yy}$ at half-fillings are themselves larger in magnitude whenever $\rho_{xx}/\rho_{yy}$ is large, implying that there is a \textit{spin-dependent} resistivity. This is qualitatively similar to the observations previously reported, and attributed tentatively to the loss of screening and hence the larger resistivity, when the topmost LL's magnetization is aligned with the net magnetization of the underlying LLs \cite{Vakili.PRL.2005, Maryenko.PRL.2015}. Note that in Figs. 2 and 3, similar to the $\rho_{xx}/\rho_{yy}$ anisotropy, the largest $\rho_{xx}$ and $\rho_{yy}$ are also observed when the 2DES is fully spin-polarized. This raises the question whether the $\rho_{xx}/\rho_{yy}$ anisotropy we observe could simply be a result of a reduction in screening and enhanced anisotropic scattering. In an ideal 2DES with zero electron layer thickness, $\rho_{xx}$ and $\rho_{yy}$ should be equal in tilted fields as $B_{||}$ should not affect the electron motion. In quasi-2D systems with finite layer thickness, $B_{||}$ couples to the out-of-plane motion of electrons and can result in anisotropic transport, with larger resistance when current flow is perpendicular to $B_{||}$ \cite{Shayegan.Article.1990, Sarma.PRL.2000}. However, this is opposite to what we observe. We conclude that the anisotropies we observe are very likely related to the formation of interaction-induced nematic phases, rather than disorder-induced, anisotropic scattering.

In summary, we report signatures of nematic phases in an AlAs 2DES. The results provide evidence that the strength and robustness of the nematic phases at half-filled LLs critically depend not only on the LL orbital index but also on the magnetization of the topmost LL relative to the magnetization of the fully-occupied, underlying LLs (Fig. 4). The data attest to the hitherto ignored role of the spin degree of freedom in stabilizing a nematic ground state of an interacting 2DES.

\begin{acknowledgments}
We acknowledge support through the National Science Foundation (Grants DMR 1709076 and ECCS 1508925) for measurements, and the National Science Foundation (Grant No. MRSEC DMR 1420541), the U.S. Department of Energy Basic Energy Science (Grant No. DE-FG02-00-ER45841), and the Gordon and Betty Moore Foundation (Grant No. GBMF4420 for sample fabrication and characterization. This research is funded in part by QuantEmX grants from Institute for Complex Adaptive Matter and the Gordon and Betty Moore Foundation through Grant No. GBMF5305 to M. S. H., M. K. M., and M. S. A portion of this work was performed at the National High Magnetic Field Laboratory, which is supported by National Science Foundation Cooperative Agreement No. DMR-1644779 and the State of Florida. We thank S. Hannahs, T. Murphy, J. Park, H. Baek, and G. Jones at NHMFL for technical support. We also thank J. K. Jain, S. A. Kivelson and Inti Sodemann for illuminating discussions.
\end{acknowledgments}


\begin{thebibliography}{99}
\bibitem{footnote2} As we discuss in \cite{Suppl.Mat.}, there are some subtle differences between the anisotropic phases seen in Figs. 2 and 3 and those reported in GaAs 2DESs.
\end{thebibliography}
\end{document}